\documentclass[lettersize,journal]{IEEEtran}
\usepackage{amsmath,amsfonts}
\usepackage{algorithmic}
\usepackage{algorithm}
\usepackage{array}
\usepackage{textcomp}
\usepackage{stfloats}
\usepackage{url}
\usepackage{verbatim}
\usepackage{graphicx}
\usepackage{cite}
\usepackage{graphicx} 
\usepackage{acronym}
\usepackage{amsmath}
\usepackage{amsfonts}
\usepackage{xcolor}
\usepackage{fancyhdr}
\usepackage{tabularray}
\usepackage{multirow}
\usepackage{booktabs}
\usepackage{amssymb}
\usepackage{url}
\usepackage[caption=false,font=footnotesize]{subfig}

\acrodef{IIoT}[IIoT]{Industrial Internet of Things}
\acrodef{CIR}[CIR]{channel impulse response}
\acrodef{DCIR}[DCIR]{Directional Channel Impulse Response}
\acrodef{RSRP}[RSRP]{Reference Signal Received Power}
\acrodef{MPC}[MPC]{multipath component}
\acrodef{DL}[DL]{Deep Learning}
\acrodef{QoS}[QoS]{Quality of Service}
\acrodef{PQoS}[PQoS]{Predictive Quality of Service}
\acrodef{DPQoS}[DPQoS]{Distributed Predictive Quality of Service}
\acrodef{LOS}[LOS]{Line-of-sight}
\acrodef{NLOS}[NLOS]{Non-line-of-sight}
\acrodef{ODE}[ODE]{Ordinary Differential Equation}
\acrodef{GLNN}[GLNN]{Graph-Liquid Neural Network}
\acrodef{RIS}[RIS]{Reflective Intelligent Surface}
\acrodef{QoS}[QoS]{Quality of Service}
\acrodef{PQoS}[PQoS]{Predictive Quality of Service}
\acrodef{DPQoS}[DPQoS]{Distributed Predictive Quality of Service}
\acrodef{ML}[ML]{Machine Learning}
\acrodef{GML}[GML]{Machine Learning on Graphs}
\acrodef{UE}[UE]{User Equipment}
\acrodef{BS}[BS]{Base Station}
\acrodef{NN}[NN]{Neural Network}
\acrodef{RNN}[RNN]{Recurrent Neural Network}
\acrodef{CNN}[CNN]{Convolutional Neural Network}
\acrodef{LTC}[LTC]{Liquid Time Constant Network}
\acrodef{MIMO}[MIMO]{Multiple-Input-Multiple-Output}
\acrodef{GNN}[GNN]{Graph Neural Network}
\acrodef{STGNN}[STGNN]{Spatio-Temporal Graph Neural Network}
\acrodef{PPP}[PPP]{Poisson Point Process}
\acrodef{LSTM}[LSTM]{Long-Short Term Memory}
\acrodef{PL}[PL]{path-loss}
\acrodef{CDF}[CDF]{cumulative distribution function}
\acrodef{RMS}[RMS]{root mean square}
\acrodef{SC}[SC]{Sparse clutter}
\acrodef{DC}[DC]{Dense clutter}
\acrodef{RX}[RX]{receiver}
\acrodef{TX}[TX]{transmitter}
\acrodef{PDP}[PDP]{power-delay profile}
\acrodef{SNR}[SNR]{signal to noise ratio}
\acrodef{CSI}[CSI]{channel state information}
\acrodef{sub-THz}[sub-THz]{sub-terahertz}
\acrodef{JSAC}[ISAC]{Integrated Sensing and Communication}
\acrodef{BW}[BW]{bandwidth}
\acrodef{FMCW}[FMCW]{frequency modulated continous wave}
\acrodef{FSPL}[FSPL]{free-space path loss}
\acrodef{RSS}[RSS]{received signal strength}

\fancypagestyle{firstpage}
{
    \fancyhf{}
    \fancyhead[C]{IEEE Antennas and Wireless Propagation Letters, vol. 24, no. 10, pp. 3669-3673, Oct. 2025, \url{https://doi.org/10.1109/LAWP.2025.3600373}}
}

\begin{document}
\bstctlcite{IEEEexample:BSTcontrol}

\title{Stochastic 3D Foliage Modeling at 80\,GHz: Experimental Validation and Ray Tracing Simulations}

\author{
    \IEEEauthorblockN{
    Jiri Blumenstein\IEEEauthorrefmark{1}, 
    Radek  Zavorka\IEEEauthorrefmark{1}, Josef Vychodil\IEEEauthorrefmark{1}, Tomas Mikulasek\IEEEauthorrefmark{1}, Jaroslaw Wojtun\IEEEauthorrefmark{3}, Jan M. Kelner\IEEEauthorrefmark{3}, Cezary Ziolkowski\IEEEauthorrefmark{3}, Rajeev Shukla\IEEEauthorrefmark{2}, Markus Hofer\IEEEauthorrefmark{4}, Thomas Zemen\IEEEauthorrefmark{4}, Christoph Mecklenbrauker\IEEEauthorrefmark{6}, Aniruddha Chandra\IEEEauthorrefmark{2}, Ales Prokes\IEEEauthorrefmark{1}} \\
    
    \IEEEauthorblockA{
        \IEEEauthorrefmark{3}Institute of Communications Systems, Military University of Technology, Warsaw, \textit{Poland};
        \IEEEauthorrefmark{2}ECE Department, National Institute of Technology, Durgapur, \textit{India};
        \IEEEauthorrefmark{4}AIT Austrian Institute 
        of Technology, Vienna, \textit{Austria};\\
        \IEEEauthorrefmark{6}Institute of Telecommunications, TU Wien, Vienna, \textit{Austria};
        \IEEEauthorrefmark{1}Brno University of Technology, Brno, \textit{Czechia}
    }
}

\maketitle
\thispagestyle{firstpage}

\begin{abstract}

A stochastic modeling methodology for 3D foliage is presented, aimed at enhancing ray-tracing simulations. The model supports adjustable stochastic geometry, density, and shape to capture variability in foliage structures. The model is validated through experimental measurements of representative vegetation. The influence of foliage density and size on path loss and RMS delay spread is analyzed to demonstrate the applicability of the model in the 80\,GHz frequency band.

\end{abstract}

\section{Introduction}

Understanding the behavior of the wireless channel has always been essential for the design and optimal performance of wireless communication systems. Numerous modeling approaches have been proposed in the literature \cite{8207426}; however, a comprehensive overview of all channel modeling techniques is beyond the scope of this letter. Instead, we focus on deterministic approaches, specifically ray-tracing-based site-specific channel modeling, which can provide explainable \ac{CSI} \cite{zemen_1}. The fidelity of the computed \ac{CSI} is, however, closely tied to the accuracy of the employed 3D environment models \cite{10133177}. This dependency becomes particularly pronounced at millimeter-wave frequencies, where the wavelength is on the order of millimeters, making even very small environmental details significant for the model fidelity. 

Nonetheless, an approach to 3D foliage modeling involves the use of hollow geometric objects as tree models. This method is employed, for example, in \cite{9411133}, where additional attenuation is applied based on the distance that rays travel through the foliage blocks. In \cite{s21124112}, trees are simulated as hollow cubical objects, where even the outer foliage shape (the envelope) is notably simplified. While such approximations may be sufficient for narrow-band signals, wide-band signals with high spatial resolution are more likely to exhibit artifacts due to the absence of signal interactions within the hollow volumes. In \cite{9262930}, the cubical volume is discretized into voxels, while in \cite{9524481}, each tree is modeled by a canopy and a trunk, both represented by concentric cylinders.

A new approach is enabled by the steadily increasing computational capabilities, which may solve the problem of dealing with very small environmental details. However, in a situation with randomly time-varying 3D objects (e.g. like foliage in a windy environment), gathering the said details poses another problem. Thus, hybrid channel models are proposed. In \cite{8963266}, two types of rays are introduced. The first type of rays are deterministic \acp{MPC}, interacting with buildings and ground reflections. The second type of rays are dynamic \acp{MPC}, which capture the statistical properties of the real \acp{MPC} interacting with foliage, humans, vehicles etc. In \cite{9569961} vegetation-scattering is modeled using a hybrid physics-based and data-driven approach. Foliage is modeled as a dielectric slab containing randomly oriented leaves (disks) and branches (cylinders), which act to scatter and attenuate the propagating field. 

In \cite{8668814}, point scatterers are distributed inside the tree envelope, which is, however, again of a simple geometric shape such as sphere or cylinder. The point scatterer, also used in \cite{zemen_2} to represent a roadside scattering object, is modeled as an infinitesimal entity. It does not have thickness or random rotations as e.g. real leafs on a tree. Also, the visualization of point-scatterers and the explainability of the ray-tracing is affected.
\begin{figure}[!t]
\centering
\includegraphics[width=2.2in]{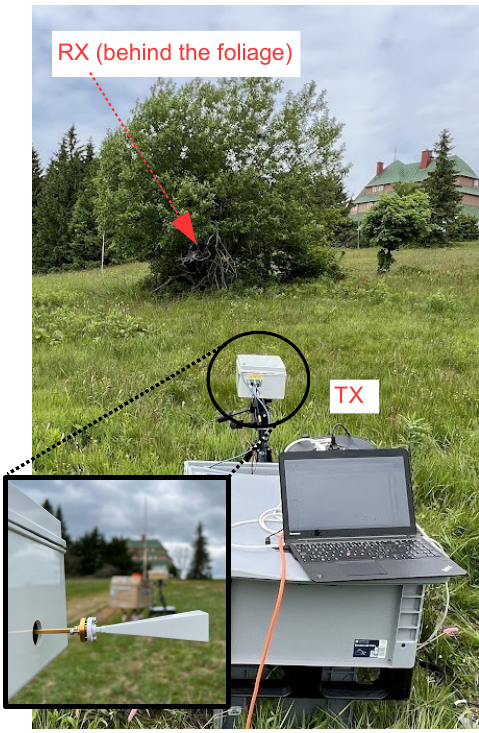}
\caption{\footnotesize{Photograph of the 80\,GHz channel sounder. The \ac{RX} is located behind the measured foliage. TX-RX separation is 30\,m with the tree 15\,m distant from the \ac{TX}.}}
\label{fig_foto}
\end{figure}
Clearly, gathering of the exact geometrical data, namely of the foliage, is deemed unpractical, if not practically impossible; however, some ray-tracing channel modeling approaches account for the presence of foliage by creating a 3D model of an exemplary tree via techniques such as 3D scanning or photogrammetry. After assigning the appropriate material properties, ray-tracing simulations are performed and the results are compared with measurement data in \cite{10366282}. The disadvantages are obviously the scanning procedure and the difficult generalization of the results. In light of the previous work mentioned above, the contribution of this paper is as follows.
\begin{itemize}
    \item A stochastic 3D foliage modeling methodology that uses several input parameters, namely: (1) the volume of the tree crown, (2) a shape deviation parameter, and (3) the density of the tree crown. The model is stochastic in the sense that, even with identical parameter settings, each realization produces a unique 3D foliage geometry. The model is validated through an extensive measurement campaign carried out outdoors with an exemplary tree. The measurements are done at 80\,GHz with a \ac{FMCW} channel sounder. A photo from the measurement campaign is shown in Fig.~\ref{fig_foto}. 
    \item We provide simulation results of \ac{PL}, \ac{RSS} and \ac{RMS} delay spread, for a wide-band 80\,GHz signal propagating through a tree under a parameter sweep.
\end{itemize}

\section{Stochastic 3D foliage model}
\begin{figure*}[!t]
\centering
\subfloat[]{\includegraphics[width=2.2in]{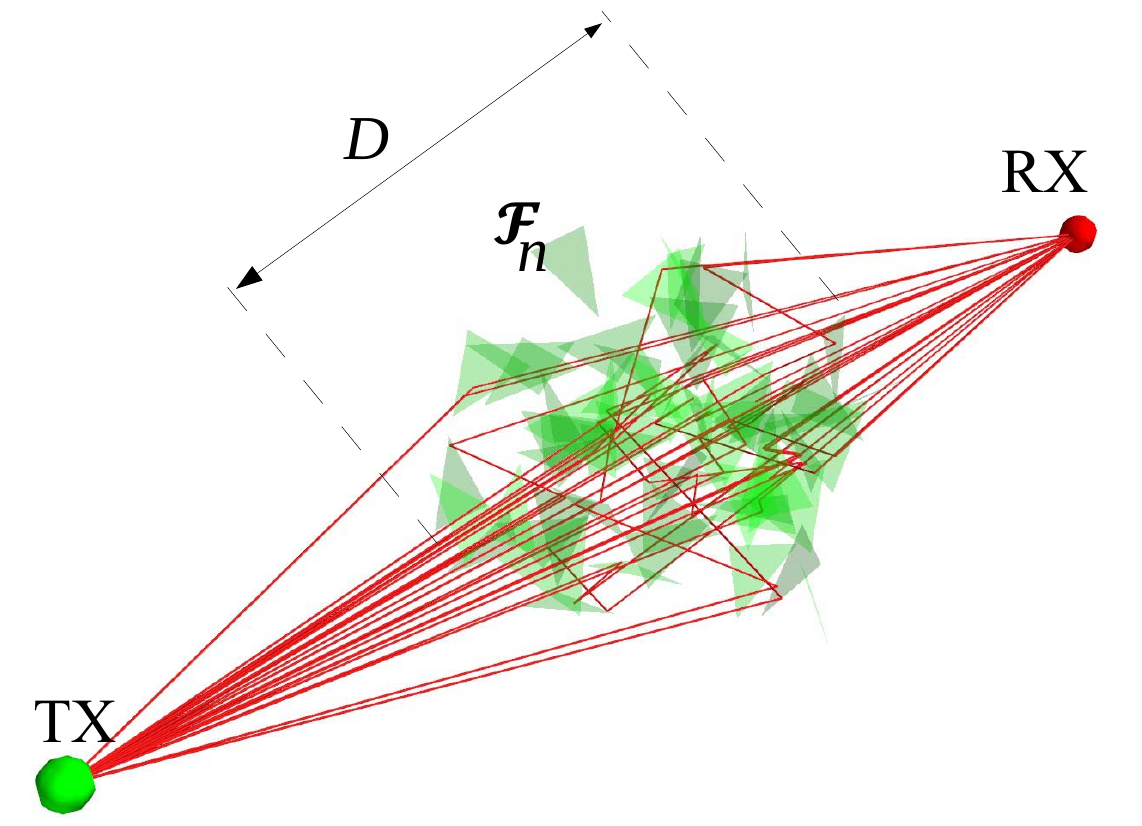}%
\label{fig_first_case}}
\hfil
\subfloat[]{\includegraphics[width=2.2in]{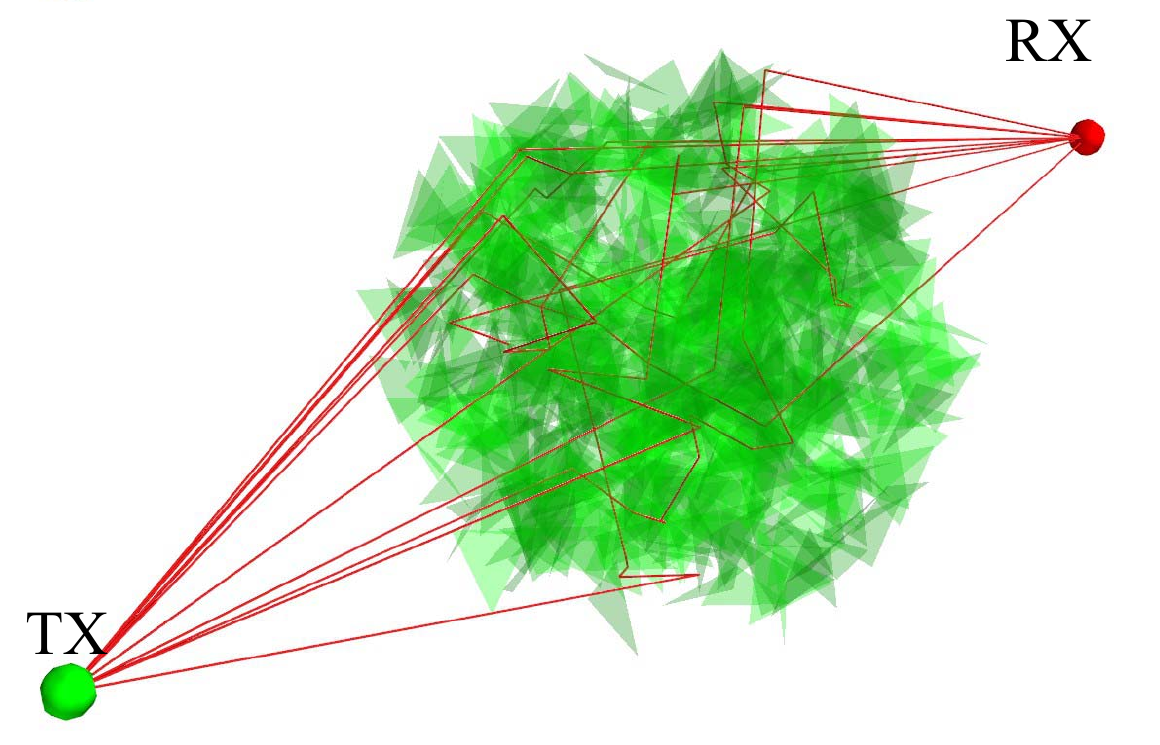}%
\label{fig_second_case}}
\hfil
\subfloat[]{\includegraphics[width=2.2in]{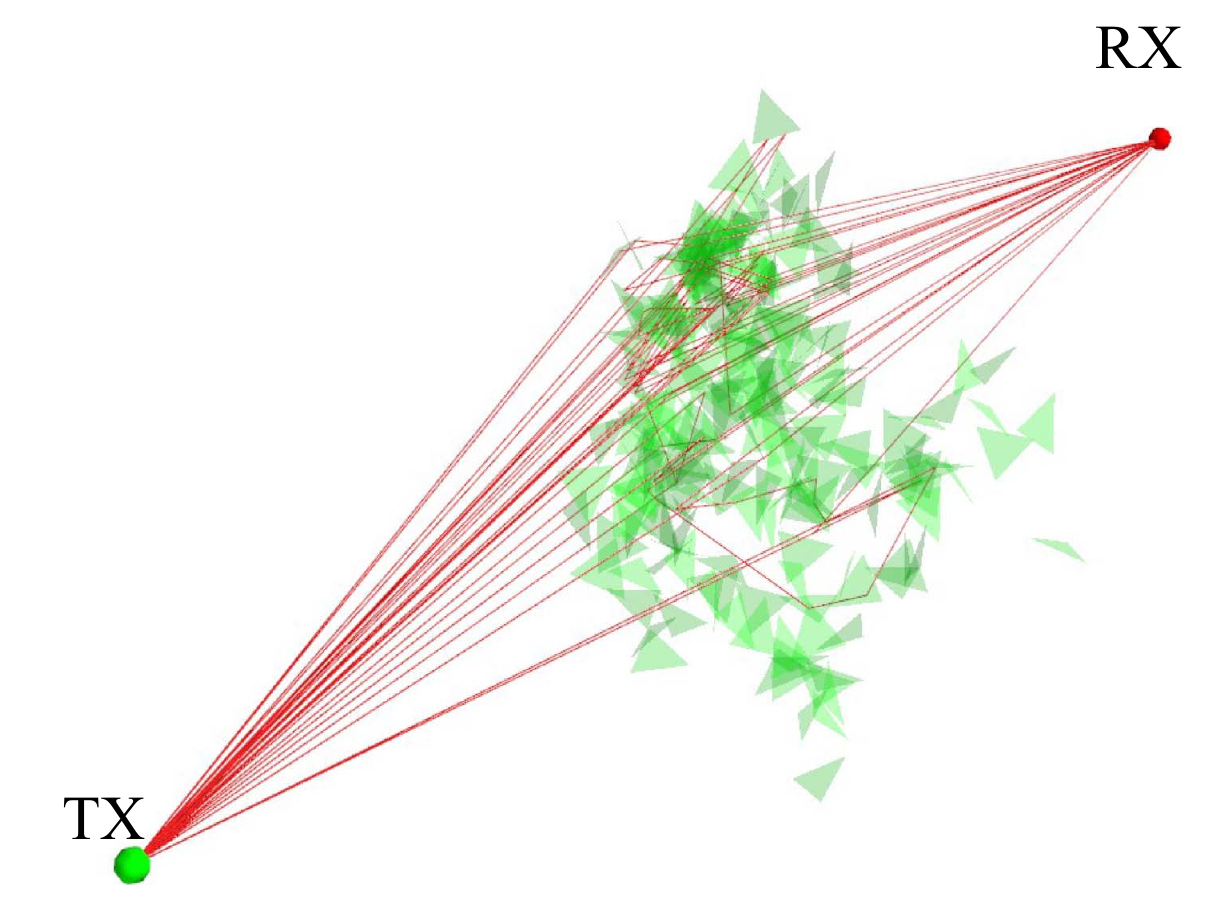}%
\label{fig_third_case}}
\caption{\footnotesize{A 3D visualization of the simulated scene with \ac{RX}, \ac{TX} and the foliage model. (a) Triangular element $\mathcal{F}_n$ represents $n$-th face and $D$ stands for diameter. This visualization assumes a density of $\rho=0.125\,\mathrm{m}^3$ and a shape perturbation parameter $\sigma=0.1$. (b) Increased density of $\rho=0.5\,\mathrm{m}^3$, $\sigma=0.1$. (c) The influence of an increased shape perturbation parameter to $\sigma=1$. Note that the internal triangle area $A=2$\,$\mathrm{m}^2$: Simulations show a low sensitivity to $A$, so the highest acceptable number was chosen to reduce the computation load. Smaller $A$ would result in higher number of internal triangles to fill~$\Omega$.}}
\label{fig_model}
\end{figure*}

This section provides a methodology for the creation of 3D foliage models. The process is divided into three main steps:
\begin{enumerate}
    \item Creation of the initial shape of the tree crown (envelope).
    \item Perturbation of the tree envelope shape with a 3D normal distribution in order to randomize the tree crown shape.
    \item Filling the perturbed volume with randomly placed and rotated triangular faces $\mathcal{F}$ with given area $A$ and density~$\rho$. A visualization of the faces $\mathcal{F}$ for different parameter settings is available in Fig. \ref{fig_model}. Note that the boundary of the perturbed tree envelope is not depicted nor used for the ray-tracing.
\end{enumerate}

\subsection{Model Creation}
The initial shape of the tree crown (envelope) $\Omega' = (\mathcal{V'}, \mathcal{F'})$ is defined by a set of vertices $\mathcal{V'} \subset \mathbb{R}^3$, and faces $\mathcal{F'} \subset \mathbb{N}^3$, and is based on a triangulated icosphere, formed by subdividing an icosahedron (although other polyhedrons with triangular faces may be used as well). Let the set of vertex positions be:
\begin{equation}
\mathcal{V'} = \{ \mathbf{v'}_1, \mathbf{v'}_2, \ldots, \mathbf{v'}_N \}, \quad \mathbf{v'}_i \in \mathbb{R}^3,
\end{equation}
where $N$ stands for the number of vertices. Each vertex lies approximately on the unit sphere, $\|\mathbf{v}_i\| \approx 1 $. The irregular shape of the tree crown $\Omega=P(\Omega')$ is perturbed by a pointwise deformation function $P : \mathbb{R}^3 \rightarrow \mathbb{R}^3$, which deforms the initial shape $\Omega'$ by adding a random displacement sampled from a 3D normal distribution. The vertices are given~as:
\begin{equation}
{\mathbf{v}}_i = \mathbf{v'}_i + \boldsymbol{\delta}_i, \quad \boldsymbol{\delta}_i \sim \mathcal{N}(\mathbf{0}, \sigma^2 \mathbf{I}).
\end{equation}
Here, $\sigma$ is a user-defined parameter controlling the perturbation strength. Let $V_0$ denote the volume of the irregularized shape, computed numerically (e.g., via tetrahedral decomposition). To achieve a target volume $V_{\text{target}}$, we scale all vertices uniformly:

\begin{equation}
s = \left( \frac{V_{\text{target}}}{V_0} \right)^{1/3}, \quad \mathbf{v}_i^{\text{final}} = s \cdot {\mathbf{v}}_i.
\end{equation}
With this scaling operation we ensure that the resulting shape $\Omega=(\{\mathbf{v}_1^{\text{final}},\mathbf{v}_2^{\text{final}},\ldots,\mathbf{v}_N^{\text{final}} \},\mathcal{F'})$ approximates the desired volume of the tree crown.

Now, the tree shape is ready to be filled with triangular faces $\mathcal{F}\subset \mathbb{N}^3$, creating an irregular and stochastic internal structure of the tree. Given a face density $\rho$ (in faces per cubic meter) and the user-defined volume $V_{\text{target}}$, the total number of internal trianges is $ Q = \left\lfloor \rho \cdot V_{\text{target}} \right\rfloor$. Let each internal triangle have area $A$, which is defined by the user. Then, assuming equilateral triangles, the side length $l = \sqrt{ \frac{4A}{\sqrt{3}} }$.
The triangular face $\mathcal{F}$ are defined in a local coordinate system centered at the origin, its vertices are:
\begin{equation}
\mathbf{p}_1 = \left(0, -\frac{l}{\sqrt{3}}, 0\right), \quad
\mathbf{p}_2 = \left(\frac{l}{2}, \frac{l}{2\sqrt{3}}, 0\right), \quad 
\end{equation}
$$
\mathbf{p}_3 = \left(-\frac{l}{2}, \frac{l}{2\sqrt{3}}, 0\right).
$$
Each triangle is randomly rotated using a Rodrigues' rotation formula with a $3 \times 3$ orthonormal matrix $R$ that satisfies:

\begin{figure*}[!t]
\centering
\subfloat[]{\includegraphics[width=3in]{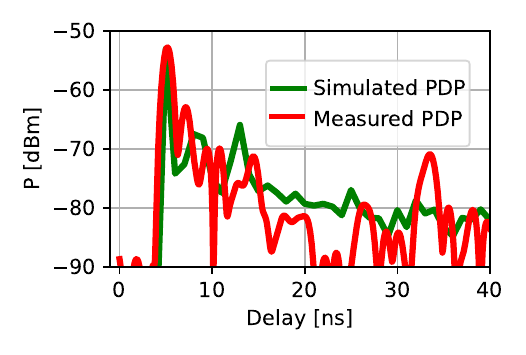}
\label{fig_pdp}}
\hfil
\subfloat[]{\includegraphics[width=3in]{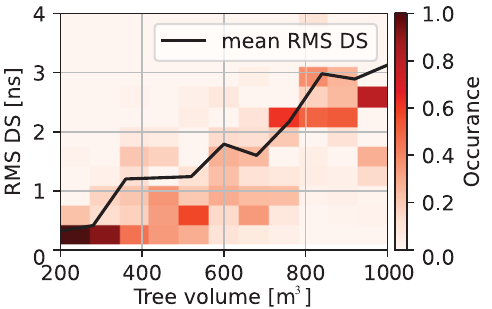}%
\label{fig_DS_volume}}
\caption{\footnotesize{(a) Measured and ray-traced PDPs (b) RMS delay spread as a function of the tree volume. We generate fifty 3D tree realizations for each simulated tree crown volume $V_\mathrm{target}\in\{200,1000\}$\,$\mathrm
{m}^3$. The depicted heatmap shows the occurrence of a certain RMS DS value.  Other parameters are: triangle density $\rho=0.125$, triangle area $A=2$\,m$^2$, perturbation strength $\sigma=0.1$\,m and number of subdivisions $n_{\text{subdiv}}=2$.}}
\end{figure*}

\begin{equation}
R^\top R = I \quad \text{and} \quad \det(R) = 1,
\end{equation}
where $^\top$ denotes a transpose operation. This ensures that $R$ performs a pure rotation without scaling. Given local triangle coordinates $\mathbf{p}_j \in \mathbb{R}^3$ (for $j = 1,2,3$), their rotated counterparts are:
\begin{equation}
\mathbf{p}_j' = R \cdot \mathbf{p}_j.
\end{equation}
The rotated triangle is then translated to a random position $\mathbf{s}_j$ utilizing a random offset $\mathbf{c}~\in~\mathbb{R}^3~\text{ such that } \mathbf{c}~\sim~\mathcal{U}(\Omega)
\quad \text{and} \quad \mathbf{c} \in \Omega$.  

Note that $\mathcal{U}(\Omega)$ denotes a uniform distribution over the region within $\Omega$, which means that every point within the volume has the same probability of being chosen. 

We write $
\mathbf{s}_j = \mathbf{c} + \mathbf{p}_j', \quad j = 1,2,3$. Then, the final vertex positions are:
\begin{equation}
\mathbf{s}_{i1} = \mathbf{c}_i + R \cdot \mathbf{p}_1, \quad
\mathbf{s}_{i2} = \mathbf{c}_i + R \cdot \mathbf{p}_2, \quad
\mathbf{s}_{i3} = \mathbf{c}_i + R \cdot \mathbf{p}_3.
\end{equation}

Repeating this approach gives $Q$ independently rotated, uniformly distributed internal triangles $\mathcal{F}$. The list of user-defiend parameters is available in Table \ref{tab:parameters}.

\renewcommand{\arraystretch}{1.3}  
\begin{table}[h!]
\caption{\footnotesize{List of user-defined parameters for irregular shape generation and internal triangle placement.}}
\centering
\begin{tabular}{|c|l|l|}
\hline
\textbf{Parameter} & \textbf{Description}\\
\hline
$V_{\text{target}}$ $[\mathrm{m}^3]$ & Target volume of the outer shape \\
\hline
$\sigma [\mathrm{m}]$ & Perturbation strength (standard deviation of noise)\\
\hline
$n_{\text{subdiv}}$ $[-]$ & Number of subdivisions of the icosphere\\
\hline  
$\rho$ [$\mathrm{triangles/m}^3$] & Internal triangle density \\
\hline
$A$ [$\mathrm{m}^2$] & Area of each internal triangle \\
\hline
$r_{\text{seed}}$ $[-]$ & Random seed (optional, for reproducibility)\\
\hline
\end{tabular}
\label{tab:parameters}
\end{table}
\vspace{-15pt}

\subsection{Ray-tracing Simulation Setup}
The wireless channel parameters are obtained by the ray-tracing. In Fig. \ref{fig_model} we see the rays depicted in red. We use the open source ray-tracer Sionna, version~$1.0.2$ \cite{sionna}. The material settings are based on ITU-R P.833 recommendation~\cite{ITU-RP833} as follows: relative permittivity $\epsilon_r=17$, conductivity $\kappa=0.05$\,S/m and scattering coefficient $\mu_s=0.50$. 

Since the ray-tracing provides unlimited \ac{BW} with \acp{MPC} being infinitely narrow, we use a sinc-filter for pulse shaping to reduce the \ac{BW} to 2\,GHz. The setting of the ray-tracing engine are listed in Table \ref{tab_sionna_parameters} together with related channel sounder parameters.

\renewcommand{\arraystretch}{1.3}  
\begin{table}[h!]
\caption{\footnotesize{List of the channel sounder parameters and  ray-tracing settings.}}
\centering
\begin{tabular}{|c|l|l|l|}
\hline
\textbf{Parameter} & \textbf{Measurement} &  \textbf{Ray-tracing} \\ \hline
        Frequency, Bandwidth & 80\,GHz, 2.048\,GHz & 80\,GHz, 2\,GHz \\ \hline
        RX and TX Antenna, Gain & Horn, 24.8\,dBi & Isotropic, 0\,dBi \\  \hline
        TX Power (w/o the ant. gain) & 34\,dBm & 0\,dBm \\  \hline
        Number of Candidate Paths & - & $2 \times 10^6$ \\  \hline
        Maximum Path Depth & - & 25 \\
\hline
\end{tabular}
\label{tab_sionna_parameters}
\end{table}
\vspace{-15pt}

\section{Verification measurement campaign}
The verification of the ray-tracing simulations with the proposed 3D foliage model is performed. The measurements are done with a \ac{BW} of 2\,GHz and horn antennas with 24.8\,dBi gain. A photo from the measurement is depicted in Fig. \ref{fig_foto}. We have used \ac{FMCW} channel sounder with a FMCW sweep duration of $T=8$\,$\mathrm{\mu s}$, where snapshot averaging is used to additionally enhance the \ac{SNR}. In \cite{10703038} we provide more details on the measurement campaign.

\subsection{Wireless Channel Parameters}
The \ac{CIR} is defined~as:
\begin{equation}
h(t, \tau) = \sum_{m=1}^M G_m a_m(t)e^{j2\pi f_ct+\psi_m} \delta(\tau - \tau_m),
\end{equation}
where $M$ is the number of \acp{MPC}, $a_m$ and $\psi_m$ are the amplitude and phase of the $m$-th \ac{MPC}, $f_\mathrm{c}$ is the carrier frequency. The term $\delta(\tau - \tau_m)$ represents the Dirac delta function, indicating the time delay of the multipath component~$m$. The effects of the transmitting and receiving antennas are integrated in the term $G_m$. 
The \ac{PDP} is given~as:
\begin{equation}
P(\tau ) = \mathbb{E}{\left| {h(t,\tau )} \right|^2}.
\end{equation} When processing the measurement data, the expectation is approximated by averaging over multiple snapshots. In contrast, in the case of the ray-tracing simulation, multiple random 3D foliage models are generated using the same parameter settings, and the resulting \ac{PDP} is obtained by averaging the corresponding~squared \acp{CIR} in absolute value.



An important parameter describing temporal dispersion of the channel is the second-order central moment of the \ac{PDP}, referred to as the \ac{RMS} delay spread~\cite{molisch2012wireless}:
\begin{equation}{\mathrm{D}_{{\text{RMS}}}} = \sqrt {\frac{{\sum_\tau {(\tau - {{\bar \tau }})^2P(\tau )} }}{{\sum_\tau {P(\tau )} }}} ,{\text{where }}\bar \tau = \frac{{\sum_\tau {\tau P(\tau )} }}{{\sum_\tau {P(\tau )} }}.
\end{equation}

Another parameter evaluated is the \ac{PL}, which is given~as 
$\mathrm{PL} = \sum_{\tau} |h(\tau)|^2$ (note that the \ac{TX} power is $0 \,\mathrm{dBm}$).

In Fig. \ref{fig_pdp} we depict the \ac{PDP} calculated from the ray-traced \acp{CIR} together with the measured \ac{PDP}. Note that the aim of this modeling method is not to exactly replicate the measured \acp{MPC}, but rather to capture second-order statistics while maintaining the stochastic nature of the modeling approach and an understandable parametrization of the model.

\begin{figure*}[!t]
\centering
\subfloat[]{\includegraphics[width=2.9in]{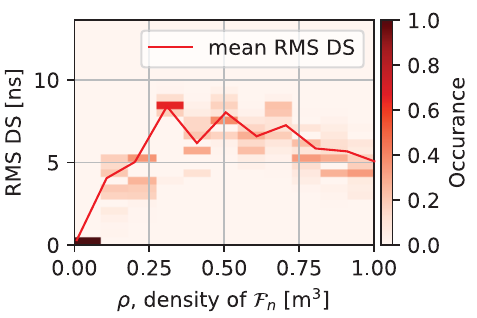}
\label{fig_DS_density}}
\hfil
\subfloat[]{\includegraphics[width=3in]{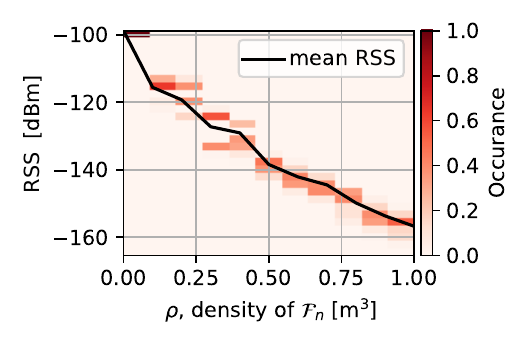}
\label{fig_PL_density}}
\caption{\footnotesize{(a) RMS delay spread and (b) Received signal strength evaluation of fifty  realizations of the tree model for each simulated leaf density $\rho$.}}
\end{figure*}

\subsection{Simulation Results}
In Fig. \ref{fig_DS_volume} we see the influence of the tree crown volume $V_\mathrm{target}\in\{200,1000\}$\,$\mathrm{m}^3$ for $A=2$\,$\mathrm{m}^2$, $\sigma=0.1$ and $n_{\text{subdiv}}=2$. For $\rho=0.125$, which is rather sparse (see Fig. \ref{fig_first_case}), the resulting $\mathrm{D}_{{\text{RMS}}}=0.5$\,ns for the smallest volume $V_\mathrm{target}=200\,\mathrm{m}^3$. As $V_\mathrm{target}$ increases, $\mathrm{D}_{{\text{RMS}}}$ increases to 3\,ns.

Additionally, the $\mathrm{D}_{{\text{RMS}}}$ is evaluated in Fig. \ref{fig_DS_density}
 as a dependence on the density $\rho$ for a given volume of the tree crown. The simulation is setup such that for each density, we generate fifty random 3D models of the foliage with otherwise constant parameters. Thus, we are able to depict a color-coded histogram together with the mean value of $\mathrm{D}_{{\text{RMS}}}$ and the $\mathrm{RSS}$. 

 We see that increasing the tree crown density from $\rho=0$ (which is free space) to $\rho=1$ leads to an increase in the RMS delay spread from $\mathrm{D}_{{\text{RMS}}}=0$\,$\mathrm{ns}$ to $\mathrm{D}_{{\text{RMS}}}=8$\,$\mathrm{ns}$. It is notable that from $\rho=0.6$ the $\mathrm{D}_{{\text{RMS}}}$ slightly decreases. The measured data exhibit $\mathrm{D}_{{\text{RMS}}}=8.5$\,$\mathrm{ns}$.
 
As for the \ac{RSS} and PL, it holds that $\mathrm{PL}=-\mathrm{RSS}$ (assuming TX power is 0\,dBm). Thus, in Fig. \ref{fig_PL_density}, we see $\mathrm{RSS}=-100$\,dBm for $\rho=0$, thus $\mathrm{PL}=100$\,dB (agrees with \ac{FSPL} at 30\,m of TX-RX separation). As the density $\rho$ increases, the losses increase as well to 60\,dB for $\rho=1$ (assuming we deduce the \ac{FSPL}). 

\begin{figure}[H]
\centering
\includegraphics[width=3in]{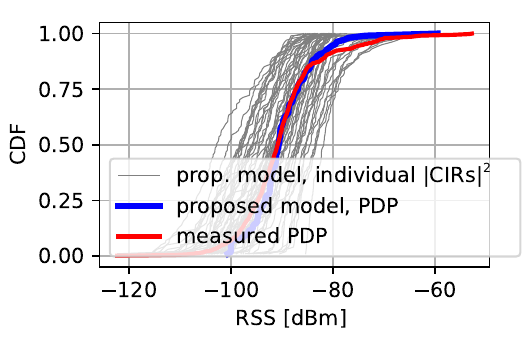}
\caption{\footnotesize{CDF evaluation of the simulated and measured PDPs as shown~in~Fig.~\ref{fig_pdp}. The CDFs of the individual CIRs $|h(t,\tau)|^2$ are depicted with varied leaf density $\rho \in \{0,1\}$.}}
\label{fig_CDF}
\end{figure}

In order to compare the measurements with the outcome of the ray-tracing simulations, we plot the \acp{CDF} of the measured \acp{PDP}, the ray-traced \acp{PDP} and the ray-traced \acp{CIR}. In total, in Fig. \ref{fig_CDF} we see fifty \acp{CIR} for each tree density value $\rho \in \{0,1\}$.

\section{Conclusion}
In this letter we show a methodology for stochastic 3D foliage modeling, intended for ray-tracing simulations. Sweeping through the model parameters, we show the resulting wireless channel parameters such as path loss and RMS delay spread. By comparing \acp{CDF}, we demonstrate a match between the measured and ray-traced data.

\section{Acknowledgments}
\footnotesize{The research described in this paper was financed by the Czech Science Foundation, Project No. 23-04304L, "Multi-band prediction of millimeter-wave propagation effects for dynamic and fixed scenarios in rugged time varying environments" and by the National Science Centre, Poland, Project No. 2021/43/I/ST7/03294, through the OPUS-22 (LAP) Call in the Weave Program.}




\bibliographystyle{IEEEtran}
\renewcommand{\IEEEbibitemsep}{0pt plus 0.5pt}
\def\baselinestretch{0.95}
\bibliography{references}

@IEEEtranBSTCTL{IEEEexample:BSTcontrol,
CTLuse_forced_etal = "yes",
CTLmax_names_forced_etal = "1",
CTLnames_show_etal="1",
}

@book{molisch2012wireless,
  title={Wireless communications},
  author={Molisch, Andreas F},
  year={2012},
  publisher={John Wiley \& Sons}
}

@INPROCEEDINGS{8963266,
  author={Yang, Songjiang and Zhang, Jiliang and Zhang, Jie},
  booktitle={2019 International Symposium on Antennas and Propagation (ISAP)}, 
  title={Impact of Foliage on Urban MmWave Wireless Propagation Channel: A Ray-tracing Based Analysis}, 
  year={2019},
  volume={},
  number={},
  pages={1-3},
  doi={}}

@INPROCEEDINGS{9411133,
  author={Taygur, Mehmet Mert and Eibert, Thomas F.},
  booktitle={2021 15th European Conference on Antennas and Propagation (EuCAP)}, 
  title={Foliage Modeling for Urban Ray-Tracing Simulations Using Satellite Images}, 
  year={2021},
  volume={},
  number={},
  pages={1-4},
  doi={10.23919/EuCAP51087.2021.9411133}}

@ARTICLE{10366282,
  author={Lai, Chiehping and Senic, Damir and Gentile, Camillo and Senic, Jelena and Golmie, Nada},
  journal={IEEE Access}, 
  title={Raytracing Digital Foliage at Millimeter-Wave: A Case Study on Calibration Against 60-{GHz} Channel Measurements on Summer and Winter Trees}, 
  year={2023},
  volume={11},
  number={},
  pages={145931-145943},
  doi={10.1109/ACCESS.2023.3345248}}

@ARTICLE{9569961,
  author={Zhang, Peize and Yi, Cheng and Yang, Bensheng and Wang, Haiming and Oestges, Claude and You, Xiaohu},
  journal={IEEE Transactions on Antennas and Propagation}, 
  title={Predictive Modeling of Millimeter-Wave Vegetation-Scattering Effect Using Hybrid Physics-Based and Data-Driven Approach}, 
  year={2022},
  volume={70},
  number={6},
  pages={4056-4068},
  doi={10.1109/TAP.2021.3118463}}

@Article{s21124112,
AUTHOR = {Rodríguez-Corbo, Fidel Alejandro and Azpilicueta, Leyre and Celaya-Echarri, Mikel and Lopez-Iturri, Peio and Alejos, Ana V. and Shubair, Raed M. and Falcone, Francisco},
TITLE = {Deterministic and Empirical Approach for Millimeter-Wave Complex Outdoor Smart Parking Solution Deployments},
JOURNAL = {Sensors},
VOLUME = {21},
YEAR = {2021},
NUMBER = {12},
ARTICLE-NUMBER = {4112},
PubMedID = {34203774},
ISSN = {1424-8220},
DOI = {10.3390/s21124112}
}

@ARTICLE{9262930,
  author={Charbonnier, Romain and Lai, Chiehping and Tenoux, Thierry and Caudill, Derek and Gougeon, Grégory and Senic, Jelena and Gentile, Camillo and Corre, Yoann and Chuang, Jack and Golmie, Nada},
  journal={IEEE Transactions on Vehicular Technology}, 
  title={Calibration of Ray-Tracing With Diffuse Scattering Against 28-{GHz} Directional Urban Channel Measurements}, 
  year={2020},
  volume={69},
  number={12},
  pages={14264-14276},
  doi={10.1109/TVT.2020.3038620}}

@ARTICLE{9524481,
  author={da Silva, Jean Carneiro and Costa, Emanoel},
  journal={IEEE Transactions on Antennas and Propagation}, 
  title={A Ray-Tracing Model for Millimeter-Wave Radio Propagation in Dense-Scatter Outdoor Environments}, 
  year={2021},
  volume={69},
  number={12},
  pages={8618-8629},
  doi={10.1109/TAP.2021.3083779}}

@ARTICLE{8668814,
  author={Leonor, Nuno R. and Fernandes, Telmo R. and García Sánchez, Manuel and Caldeirinha, Rafael F. S.},
  journal={IEEE Transactions on Antennas and Propagation}, 
  title={A {3-D} Model for Millimeter-Wave Propagation Through Vegetation Media Using Ray-Tracing}, 
  year={2019},
  volume={67},
  number={6},
  pages={4313-4318},
  doi={10.1109/TAP.2019.2905957}}

@ARTICLE{8207426,
  author={Hemadeh, Ibrahim A. and Satyanarayana, Katla and El-Hajjar, Mohammed and Hanzo, Lajos},
  journal={IEEE Communications Surveys \& Tutorials}, 
  title={Millimeter-Wave Communications: Physical Channel Models, Design Considerations, Antenna Constructions, and Link-Budget}, 
  year={2018},
  volume={20},
  number={2},
  pages={870-913},
  doi={10.1109/COMST.2017.2783541}}

@INPROCEEDINGS{10133177,
  author={Kleijer, Marjolijn and Steinböck, Gerhard and Olsson, Bengt-Erik and Johansson, Martin and Smolders, Bart},
  booktitle={2023 17th European Conference on Antennas and Propagation (EuCAP)}, 
  title={Impact of Facade Details on Radio Propagation at 28 {GHz}}, 
  year={2023},
  volume={},
  number={},
  pages={1-5},
  doi={10.23919/EuCAP57121.2023.10133177}}

@ARTICLE{10703038,
  author={Zavorka, Radek and Mikulasek, Tomas and Vychodil, Josef and Blumenstein, Jiri and Chandra, Aniruddha and Hammoud, Hussein and Kelner, Jan M. and Ziółkowski, Cezary and Zemen, Thomas and Mecklenbräuker, Christoph and Prokes, Ales},
  journal={IEEE Access}, 
  title={Characterizing the 80 {GHz} Channel in Static Scenarios: Diffuse Reflection, Scattering, and Transmission Through Trees Under Varying Weather Conditions}, 
  year={2024},
  volume={12},
  number={},
  pages={144738-144749},
  doi={10.1109/ACCESS.2024.3472003}}

@software{sionna,
 title = {Sionna},
 author = {Hoydis, Jakob and Cammerer, Sebastian and {Ait Aoudia}, Fayçal and
 Nimier-David, Merlin and Maggi, Lorenzo and Marcus, Guillermo and Vem, Avinash and Keller,
 Alexander},
 note = {https://nvlabs.github.io/sionna/},
 year = {2022},
 version = {1.0.2}
}

@techreport{ITU-RP833,
  author      = {{International Telecommunication Union}},
  title       = {{Attenuation in vegetation}},
  institution = {ITU Radiocommunication Sector},
  number      = {Recommendation ITU-R P.833-9},
  year        = {2016},
  note        = {\url{https://www.itu.int/rec/R-REC-P.833-9-201609-I/en}},
}

@ARTICLE{zemen_1,
  author={Zemen, Thomas and Gomez-Ponce, Jorge and Chandra, Aniruddha and Walter, Michael and Aksoy, Enes and He, Ruisi and Matolak, David and Kim, Minseok and Takada, Jun-ichi and Salous, Sana and Valenzuela, Reinaldo and Molisch, Andreas F.},
  journal={IEEE Communications Magazine}, 
  title={Site-Specific Radio Channel Representation for {5G} and {6G}}, 
  year={2024},
  volume={},
  number={},
  pages={1-8},
  doi={10.1109/MCOM.001.2400355}}

@ARTICLE{zemen_2,
  author={Karedal, Johan and Tufvesson, Fredrik and Czink, Nicolai and Paier, Alexander and Dumard, Charlotte and Zemen, Thomas and Mecklenbrauker, Christoph F. and Molisch, Andreas F.},
  journal={IEEE Transactions on Wireless Communications}, 
  title={A geometry-based stochastic {MIMO} model for vehicle-to-vehicle communications}, 
  year={2009},
  volume={8},
  number={7},
  pages={3646-3657},
  doi={10.1109/TWC.2009.080753}}

\end{document}